\documentclass[preprint2]{aastex}
\usepackage{rotating}
\usepackage{graphicx}
\DeclareGraphicsExtensions{.ps} 

\shorttitle{\indent \def Multi-episode chromospheric evaporation} \shortauthors{Tian et al.}

\begin{document}

\title{Multi-episode chromospheric evaporation observed in a solar flare}

\author{H. Tian\altaffilmark{1}, N.-H. Chen\altaffilmark{2}} 
\altaffiltext{1}{School of Earth and Space Sciences, Peking University, 100871 Beijing, China; huitian@pku.edu.cn}
\altaffiltext{2}{Korea Astronomy and Space Science Institute, Daejeon, Republic of Korea} 

\begin{abstract}
With observations of the Interface Region Imaging Spectrograph (IRIS), we study chromospheric heating and evaporation during an M1.6 flare SOL2015-03-12T11:50. At the flare ribbons, the Mg~{\sc{ii}}~2791.59 \AA{} line shows quasi-periodic short-duration red-wing enhancement, which is likely related to repetitive chromospheric condensation as a result of episodic heating. On the contrary, the Si~{\sc{iv}}~1402.77 \AA{} line reveals a persistent red-wing asymmetry in both the impulsive and decay phases, suggesting that this line responds to both cooling downflows and chromospheric condensation. The first two episodes of red-wing enhancement occurred around 11:42 UT and 11:45 UT, when two moving brightenings indicative of heating fronts crossed the IRIS slit. The greatly enhanced red wings of the Si~{\sc{iv}}~and Mg~{\sc{ii}}~lines at these occasions are accompanied by an obvious increase in the line intensities and the HXR flux, suggesting two episodes of energy injection into the lower atmosphere in the form of nonthermal electrons. The Mg~{\sc{ii}}~k/h ratio has a small value of $\sim$1.2 at the ribbons and decreases to $\sim$1.1 at these two occasions. Correspondingly, the Fe~{\sc{xxi}}~1354 \AA{}~line reveals two episodes of chromospheric evaporation, which is characterized as a smooth decrease of the blue shift from $\sim$300 km~s$^{-1}$ to nearly zero within $\sim$3 minutes. The Fe~{\sc{xxi}}~1354 \AA{}~line is entirely blueshifted in the first episode, while appears to contain a nearly stationary component and a blueshifted component in the second episode. More episodes of blueshifted Fe~{\sc{xxi}}~emission is found around the northern ribbon in the decay phase, though no obvious response is detected in the Si~{\sc{iv}}~and Mg~{\sc{ii}}~emission. We have also examined the Fe~{\sc{xxi}} emission at the flare loop top and identified a secondary component with a $\sim$200 km~s$^{-1}$ red shift, which possibly results from the downward moving reconnection outflow. Our analysis also suggests a reference wavelength of 1354.0878$\pm$0.0072 \AA{} for this Fe~{\sc{xxi}} line.
\end{abstract}

\keywords{Sun: flares---Sun: chromosphere---Sun: transition region---line: profiles---magnetic reconnection}

\section{Introduction}
During solar flares, a large amount of energy is suddenly released through magnetic reconnection, which may heat the coronal plasma and accelerate
charged particles. The released energy is then transported downwards to the lower atmosphere in the form of thermal conduction or nonthermal electrons, resulting in significant heating of the chromosphere to temperatures up to tens of MK. Such a process may lead to an overpressure in the chromosphere, which drives the heated plasma upward to fill flare loops. This so-called chromospheric evaporation may proceed explosively or gently \citep[e.g.,][]{Fisher1985b,Gan1994,Milligan2006a,Milligan2006b,Druett2017}. In the case of explosive evaporation, the heated plasma rapidly expands upward up to several hundred km~s$^{-1}$ and downward at a speed of few tens km~s$^{-1}$ simultaneously. If the energy flux is lower than $\sim$10$^{10}$ erg cm$^{-2}$ s$^{-1}$ \citep{Fisher1985a}, heating in the chromosphere often leads to gentle evaporation upflows in a wide range of temperatures at a speed of several tens km~s$^{-1}$ with no accompanied downflows. A signifciant fraction of the soft X-Ray (SXR) and extreme ultraviolet (EUV) emission during solar flares may be produced through chromospheric evaporation.

Chromospheric evaporation has been rarely imaged from observations \citep{Liu2006,Ning2009,Nitta2012,Zhang2013}. Instead, they are usually observed through spectroscopic observations of SXR and EUV spectral lines. Prior to the launch of IRIS, spectroscopic observations of coronal emission lines at flare ribbons usually revealed blue-wing enhancement or a blueshifted component besides a nearly stationary component  \citep{Doschek1980,Feldman1980,Antonucci1982,Mason1986,Mariska1993,Czaykowska1999,Brosius2003,Harra2005,DelZanna2006,Falchi2006,Milligan2006a,Milligan2006b,Teriaca2003,Teriaca2006,Raftery2009,Milligan2009,Watanabe2010,DelZanna2011,Li2011,Graham2011,Young2013}. Entirely blueshifted profiles of these hot spectral lines were identified only in a few flares \citep{Brosius2013, Doschek2013}. Such a result implies that normally the spatial resolution of these spectrometers is too low to separate evaporation flows from the ambient plasma. The blue shift appears to increase with the line formation temperature, and it can reach up to $\sim$400~km~s$^{-1}$ at temperatures higher than 10 MK. In explosive evaporation, emission lines formed at typical transition region (TR) temperatures are often found to be redshifted. The temperature at which the Doppler shift transits from red to blue is often found in the temperature range of 0.8--2 MK \citep[e.g.,][]{Milligan2009,Chen2010,Young2013}, much higher than that predicted by most flare models \citep[e.g.,][]{Fisher1985a}. However, \cite{Liu2009} has demonstrated that the high transition temperature may be explained by continuous energy deposition throughout the impulsive phase. Strong redshifts of lines formed at 1--2 MK may also be expected when thermal conduction is suppressed \citep{Imada2015}. 

Launched around the solar maximum, the Interface Region Imaging Spectrograph \citep[IRIS,][]{DePontieu2014} has observed hundreds of solar flares. With unprecedented high spatial ($\sim$0.33$^{\prime\prime}$), spectral ($\sim$3 km~s$^{-1}$) and temporal (a few seconds) resolutions, IRIS observations have provided significant insight into the process of chromospheric evaporation. In IRIS observations, profiles of the Fe~{\sc{xxi}}~1354 \AA{} line (formed at a temperature of $\sim$11 MK) are usually found to be entirely blueshifted by up to $\sim$350~km~s$^{-1}$ at flare ribbons \citep{Tian2014,Tian2015,Young2015,Li2015,Brosius2015,Polito2015,Polito2016,Sadykov2015,Sadykov2016,Graham2015,Battaglia2015,Zhang2016b,Dudik2016,Lee2017}. Such a result is consistent with hydrodynamic simulations of a single flare loop \cite[e.g.,][]{Emslie1987,Liu2009} and suggests that the resolution of IRIS is high enough to resolve the evaporation flows or flare kernels. The blue shift of the Fe~{\sc{xxi}}~line peaks before the SXR peak and correlates with the hard X-Ray (HXR) emission in several flares, providing strong support to the scenario of chromospheric evaporation driven by nonthermal electrons \citep{Tian2015,LiD2015b,LiD2017a}. Through sit-and-stare observations, it has been found that the blue shift of Fe~{\sc{xxi}}~line often smoothly decreases exponentially from the maximum value to nearly zero within a few minutes \citep[e.g.,][]{Tian2015,Graham2015,LiD2015b}. However, cool TR lines such as Si~{\sc{iv}}~1402.77 \AA{}~are often not entirely redshifted but showing red-wing enhancement in both impulsive and decay phases. These long-lasting red shifts appear to be in contradiction with one-dimensional hydrodynamic models, which usually predict downflows lasting for only $\sim$1 min in response to impulsive heating \citep[e.g.,][]{Fisher1989,RubiodaCosta2015}. Various mechanisms, such as energy release on many strands within the flare loop \citep{Warren2016,Reep2016}, cooling downflows \citep{Tian2015}, and a long-duration energy deposition \citep{Li2017}, have been proposed to explain these persistent downflows. The parameters of these TR lines, especially the intensity and Doppler shift, are sometimes found to vary quasi-periodically on time scales of 0.5--6 minutes at flare ribbons. This quasi-periodicity is often interpreted as quasi-periodic energy injection into the chromosphere as a result of bursty reconnection \citep{LiD2015a,LiT2015,Brosius2015,Brosius2016,Zhang2016a}, though \cite{Brannon2015} attributed it to motion of field lines caused by some forms of wave in the coronal current sheet such as a tearing mode or Kelvin-Helmholtz instability. It is worth noting that in these observations the Fe~{\sc{xxi}}~line showed no quasi-periodic signature or even no emission. 

Here we report results from IRIS sit-and-stare observations of an M1.6 flare peaked around 11:50 UT on 2015 March 12. Busty enhancement of the intensity and red shift can be clearly identified at the flare ribbons for several chromospheric and TR lines. In addition, the Fe~{\sc{xxi}}~line also reveals multiple episodes of emission at one location on a ribbon. In each episode the Fe~{\sc{xxi}}~line experiences a smooth decrease of blue shift from 100--300~km~s$^{-1}$ to almost zero. 

\section{Observations and data reduction}

\begin{figure*}
\centering {\includegraphics[width=\textwidth]{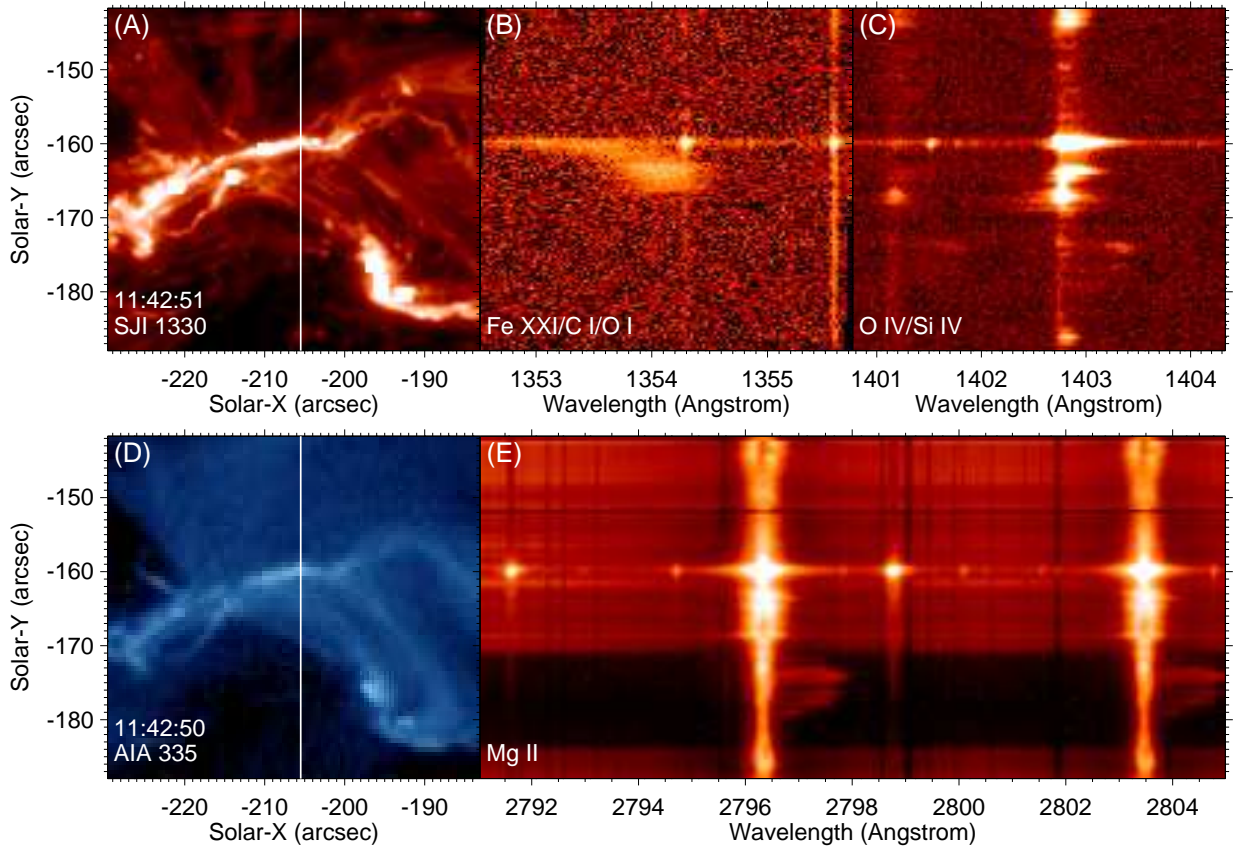}} \caption{ (A): An IRIS/SJI~1330 \AA{}~image taken at 11:42:51 UT on 2015 March 12.
(B), (C) \& (E): IRIS detector images taken at the same time in three spectral windows. (D) An SDO/AIA 335 \AA{}~image taken at 11:42:50 UT. In panels (A) and (D), the slit location is indicated by the white vertical line. A movie (m1.mp4) is available online. } \label{fig.1}
\end{figure*}

This M1.6 flare peaked around 11:50 UT on 2015 March 12 (SOL2015-03-12T11:50) in NOAA active region (AR) 12297. IRIS performed a sit-and-stare observation of this flare, with a cadence of $\sim$5.2 seconds for the spectra and $\sim$15.7 seconds for each of the 1400 \AA{}, 1330 \AA{} and 2796 \AA{} slit-jaw images (SJI) filters. The pointing of the IRIS telescope was (--235$^{\prime\prime}$, --190$^{\prime\prime}$). Due to the spectral and spatial summing, the spatial pixel size is 0.33$^{\prime\prime}$, and the spectral pixels are $\sim$0.026  \AA{} and $\sim$0.051  \AA{} for the far and near ultraviolet (FUV and NUV) wavelength bands, respectively. We used the level 2 data, where the effects of dark current, flat field, geometrical and orbital variations have all been removed. Absolute wavelength calibration has been performed by assuming zero Doppler shifts for the average profiles of several relatively strong lines from neutrals, e.g., C~{\sc{i}}~1354.288 \AA{}, O~{\sc{i}}~1355.5977 \AA{}, S~{\sc{i}}~1401.5136 \AA{}~and Mn~{\sc{i}}~2801.908 \AA{}, in a relatively quiet region to the north of the flare site (solar-Y $\geqslant$ --155$^{\prime\prime}$). This assumption can be justified since these cold lines usually have very small intrinsic velocities in the quiet solar atmosphere. Images obtained in different SJI filters and spectral windows are coaligned by examining the positions of the fiducial marks on the slit. Figure~\ref{fig.1}(A)-(C) and (E) show the IRIS SJI 1330 \AA{}~image and spectral images in three wavelength windows at 11:42:51 UT. The time sequences of these images are presented in an online video. This dataset has been previously analyzed by \cite{Tian2016} and \cite{Brannon2016} to study global sausage oscillations and draining downflows in the flare loops, respectively. 

The 335 \AA{}~images obtained by the Atmospheric Imaging Assembly \citep[AIA,][]{Lemen2012} on board the Solar Dynamics Observatory
\citep[SDO,][]{Pesnell2012} are also used here to reveal the morphology of the flare loops (see Figure~\ref{fig.1}(D) and the online video). During flares this passband samples emission mainly from the Fe~{\sc{xvi}}~335.41 \AA{}~line \citep[e.g.,][]{ODwyer2010}. The formation temperature of this line is about 3 MK. The cadence of the AIA 335 \AA{}~images is 12 seconds. The coalignment between AIA images and IRIS images were achieved by matching commonly observed features in the AIA 1600 \AA{}~(mainly FUV continuum and C~{\sc{iv}}) and IRIS 1330 \AA{}~(mainly FUV continuum and C~{\sc{ii}}) images. The AIA 335 \AA{}~images were then automatically aligned with the IRIS images since AIA images in different passbands are automatically coaligned after applying the standard data reduction routine aia\_prep.pro, which is available in SolarSoft (SSW). For AIA observations in other passbands we refer to \cite{Tian2016}.

The Reuven Ramaty High Energy Solar Spectroscopic Imager \citep[RHESSI,][]{Lin2002} was in orbit night during the occurrence of this flare. Fortunately, the Gamma-ray Burst Monitor (GBM) onboad the Fermi Gamma-ray Space Telescope observed clear enhancement of HXR in the energy range of 26-50 keV during this flare.

\section{Flare ribbons}
\subsection{General behavior at the flare ribbons}

\begin{figure*}
\centering {\includegraphics[width=\textwidth]{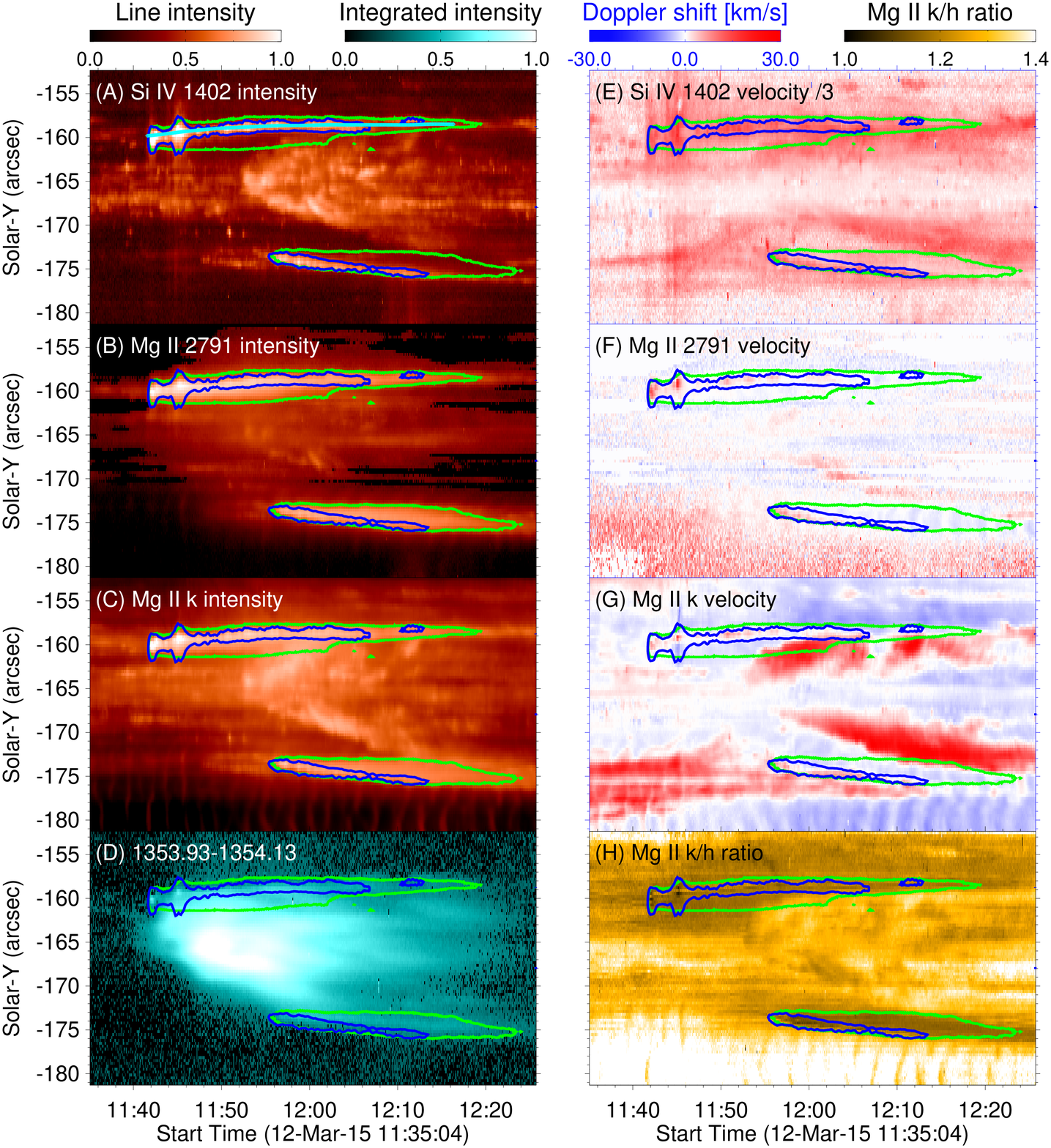}} \caption{ Temporal evolution of the line intensities (normalized to the maximum intensity in each image) and Doppler shifts (1st order moment) of the Si~{\sc{iv}}~1402.77 \AA{}, Mg~{\sc{ii}}~2791.59 \AA{} and Mg~{\sc{ii}}~k lines, the intensity integrated over the wavelength range of 1353.93 \AA{}--1354.13 \AA{} and the Mg~{\sc{ii}}~k/h ratio along the slit. The Doppler shift values of the Si~{\sc{iv}}~line have been divided by three. The green and blue contours mark the flare ribbons using intensity threshholds of Mg~{\sc{ii}}~2791.59 \AA{}~and Si~{\sc{iv}}~1402.77 \AA{}, respectively. The cyan line in panel (A) indicates the times and spatial locations at which the spectra of several emission lines are shown in Figures~\ref{fig.5}. } \label{fig.2}
\end{figure*}

\begin{figure*}
\centering {\includegraphics[width=0.8\textwidth]{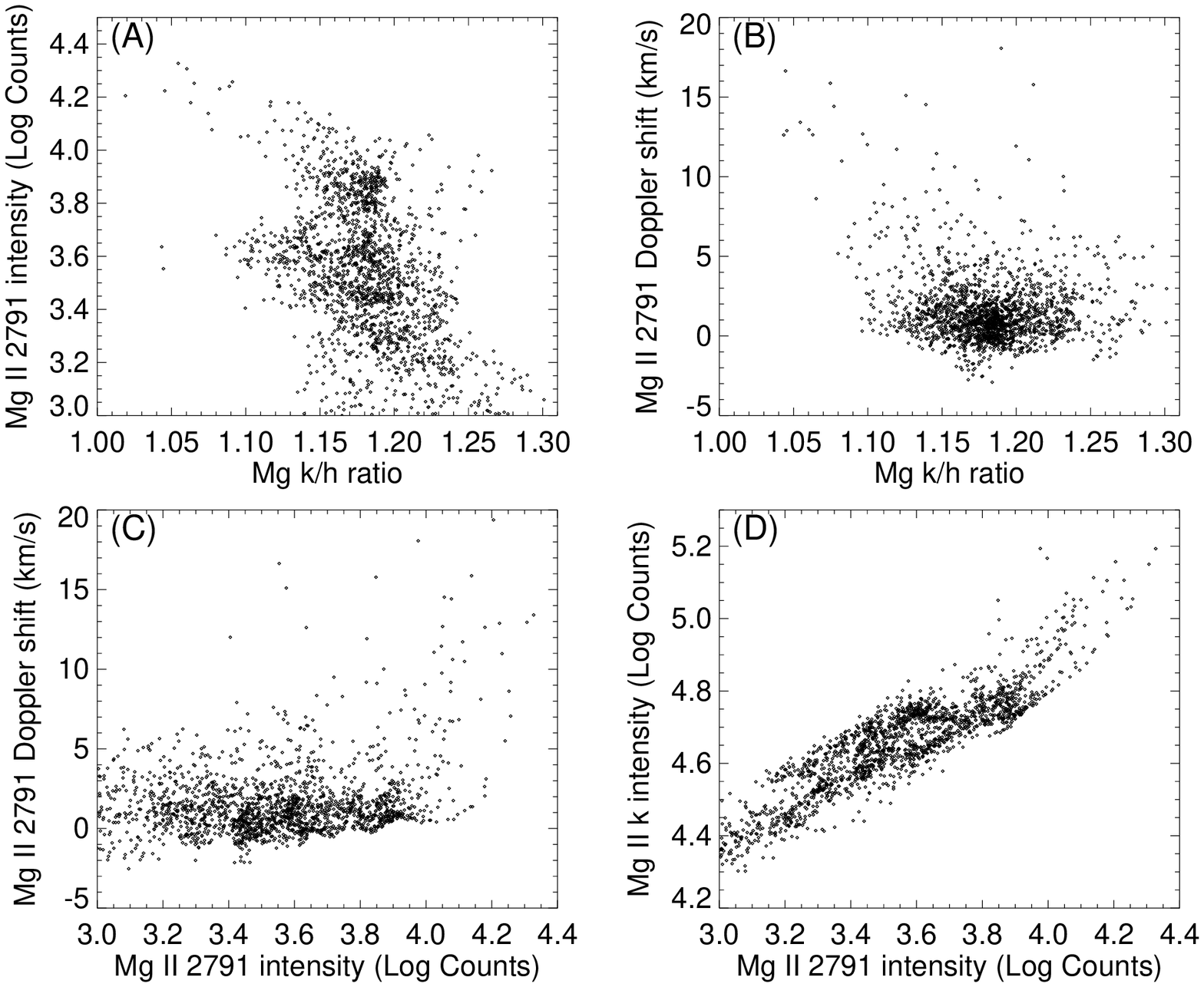}} \caption{ Scatter plots showing the relationship between different parameters within the blue contours shown in Figure~\ref{fig.2}. } \label{fig.3}
\end{figure*}

We mainly analyze the following spectral lines at the flare ribbons: Si~{\sc{iv}}~1402.77 \AA{}, Mg~{\sc{ii}}~2791.59 \AA{}, Mg~{\sc{ii}}~k~2796.35 \AA{}~and Mg~{\sc{ii}}~h~2803.52 \AA{}. The strong Si~{\sc{iv}}, Mg~{\sc{ii}}~k and h lines are core lines of IRIS, and have been demonstrated to be very sensitive to energy release during flares. Observations of IRIS often reveal the presence of many emission lines from neutrals or singly ionized ions at flare ribbons, some of which can also be seen from Figure~\ref{fig.1}. For the identification of these lines, we refer to \cite{Tian2015} and \cite{Tian2017}. Among them, the Mg~{\sc{ii}}~triplet lines at 2791.59 \AA{}, 2798.75 \AA{} and 2798.82 \AA{}~are of particular interest due to their sensitivity to heating of the lower chromosphere \citep{Pereira2015}. The latter two lines usually show up as one blended line in the IRIS data. 

The Si~{\sc{iv}}~line often reveals strong enhancement of the red wing at flare ribbons. Although the Mg~{\sc{ii}}~k, h and triplet lines become single-peak emission features at flare ribbons, they are still optically thick \citep[e.g.,][]{Kerr2015,Tian2015,Graham2015,Lee2017}. So it might be inappropriate to apply a single Gaussian fit to these line profiles. Instead, we derive the line intensity and centroid by simply integrating a spectral line profile and calculating the first order moment of the spectral line profile, respectively. The wavelength ranges used for this calculation are 1402.297 \AA{}--1403.238 \AA{}, 2791.137 \AA{}--2792.207 \AA{}, 2795.109 \AA{}--2797.655 \AA{} and 2802.238 \AA{}--2804.784 \AA{} for the Si~{\sc{iv}}~1402.77 \AA{}, Mg~{\sc{ii}}~2791.59 \AA{}, Mg~{\sc{ii}}~k~2796.35 \AA{}~and Mg~{\sc{ii}}~h~2803.52 \AA{} lines, respectively. The Doppler shift can then be derived by taking the difference between the centroid and the rest wavelength of the corresponding spectral line. A similar method was also adopted by \cite{Brosius2015} and \cite{Zhang2016a}. In Figure~\ref{fig.2} we present the temporal evolution of the intensities and Doppler shifts of the Si~{\sc{iv}}, Mg~{\sc{ii}}~2791.59 \AA{} and Mg~{\sc{ii}}~k lines, the intensity integrated in the wavelength range of 1353.93 \AA{}--1354.13 \AA{} (mainly Fe~{\sc{xxi}}~1354 \AA{}~and several cold lines) and the Mg~{\sc{ii}}~k/h ratio along the slit. The behavior of the Mg~{\sc{ii}}~h line is similar to that of Mg~{\sc{ii}}~k, and thus is not presented here.

Two flare ribbons clearly show up in the intensity images of the Si~{\sc{iv}}~and Mg~{\sc{ii}}~lines. The ribbons generally reveal red shifts of these lines. Both ribbons drift slightly along the slit during the course of the flare. The intensities and Doppler shifts of these lines appear to show quasi-periodic fluctuations, especially at the northern ribbon.

The integrated intensity of 1353.93 \AA{}--1354.13 \AA{}~is much stronger between the two ribbons than at the ribbons, indicating that the strong Fe~{\sc{xxi}}~emission mainly comes from the flare loop apexes and legs. At locations of the flare loops, we also see prominent emission from the Si~{\sc{iv}}~and Mg~{\sc{ii}}~k lines, which appears to be associated with significant red shift. These large red shifts are associated with the "C"-shape spectral feature visible in the strong chromospheric and TR lines (see the online video). They obviously result from cooling of the heated plasma in the flare loops, as discussed by \cite{Brannon2016}.  

In the optically thin case, the Mg~{\sc{ii}}~k/h ratio should be 2. Our observation shows that the Mg~{\sc{ii}}~k/h ratio is far less than 2 at the flare ribbons. Figures~\ref{fig.2}(H) reveals a ratio of $\sim$1.5 in the sunspot region to the south of the southern ribbon. While in other regions the ratio is generally less than 1.4. Interestingly, the values of Mg~{\sc{ii}}~k/h ratio at the two bright flare ribbons appear to be the smallest, mostly in the range of 1.1--1.2.  

Within the ribbons, there also appears to be a negative correlation between Mg~{\sc{ii}}~k/h ratio and Mg~{\sc{ii}}~2791.59 \AA{}~intensity (Figure~\ref{fig.3}(A)). The scatter plots shown in Figure~\ref{fig.3}(B) \& (C) reveal no obvious correlation, though there is a weak trend that a larger intensity of Mg~{\sc{ii}}~2791.59 \AA{}~is associated with a stronger red shift of the same line and a smaller Mg~{\sc{ii}}~k/h ratio. The intensities of the Mg~{\sc{ii}}~triplet lines and the two Mg~{\sc{ii}}~resonant lines are positively correlated at the ribbons, which is evident from Figure~\ref{fig.3}(D).

\subsection{Multi-episode evaporation at the northern ribbon}

\begin{figure*}
\centering {\includegraphics[width=\textwidth]{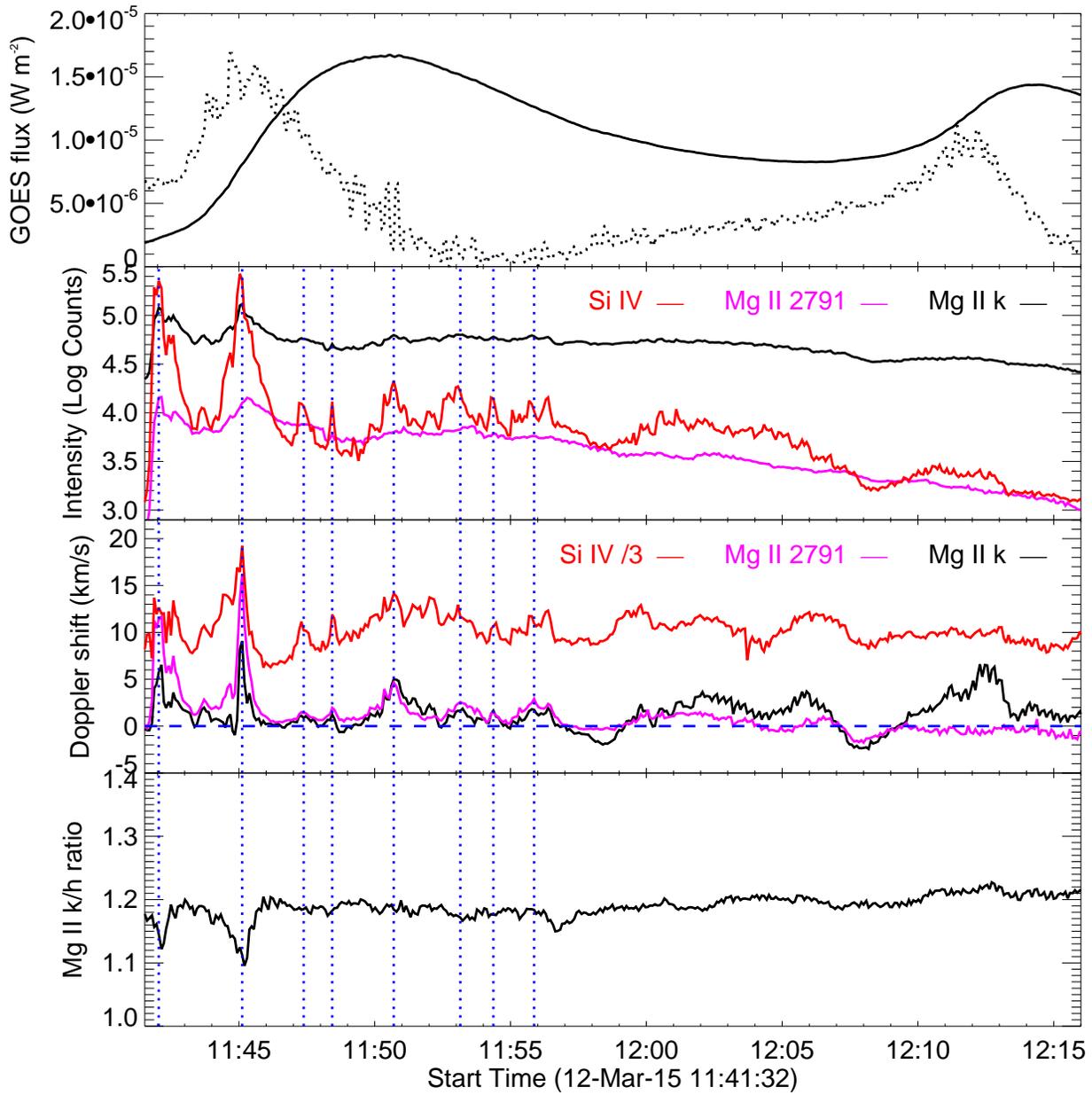}} \caption{ Upper panel: GOES 1-8 \AA{} flux (solid line) and its time derivative (dotted line). Note that the peak after 12:00 UT is produced by another flare that is not studied here. Lower three panels: Temporal evolution of the intensities and Doppler shifts of Si~{\sc{iv}}~1402.77 \AA{}, Mg~{\sc{ii}}~2791.59 \AA{} and Mg~{\sc{ii}}~k, and the Mg~{\sc{ii}}~k/h ratio, at the northern ribbon. The Doppler shift values of the Si~{\sc{iv}}~line have been divided by three. The vertical dotted lines indicate some instants when the red shifts of these lines are enhanced. } \label{fig.4}
\end{figure*}

\begin{figure*}
\centering {\includegraphics[width=\textwidth]{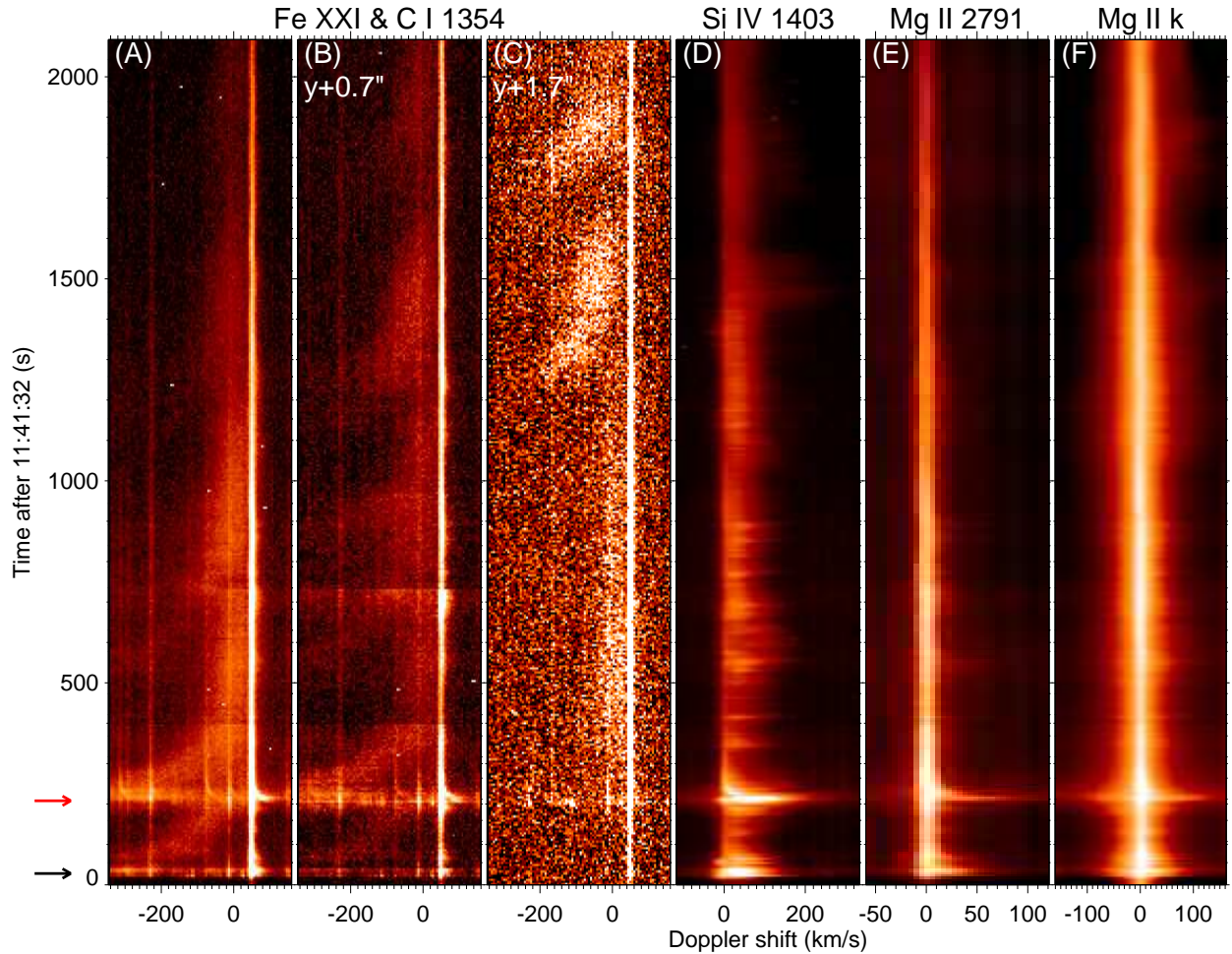}} \caption{ Temporal evolution of the IRIS spectra in four spectral windows at the northern ribbon. (A), (D)-(F) Line profiles of Fe~{\sc{xxi}}~1354 \AA{}, Si~{\sc{iv}}~1402.77 \AA{}, Mg~{\sc{ii}}~2791.59 \AA{} and Mg~{\sc{ii}}~k along the cyan line shown in Figure~\ref{fig.2}(A). (B)-(C) Same as (A) but along two lines that are 0.7$^{\prime\prime}$ and 1.7$^{\prime\prime}$ above the cyan line, respectively. The black and red arrows indicate the times of 11:42 UT and 11:45 UT, respectively, } \label{fig.5}
\end{figure*}

\begin{figure*}
\centering {\includegraphics[width=\textwidth]{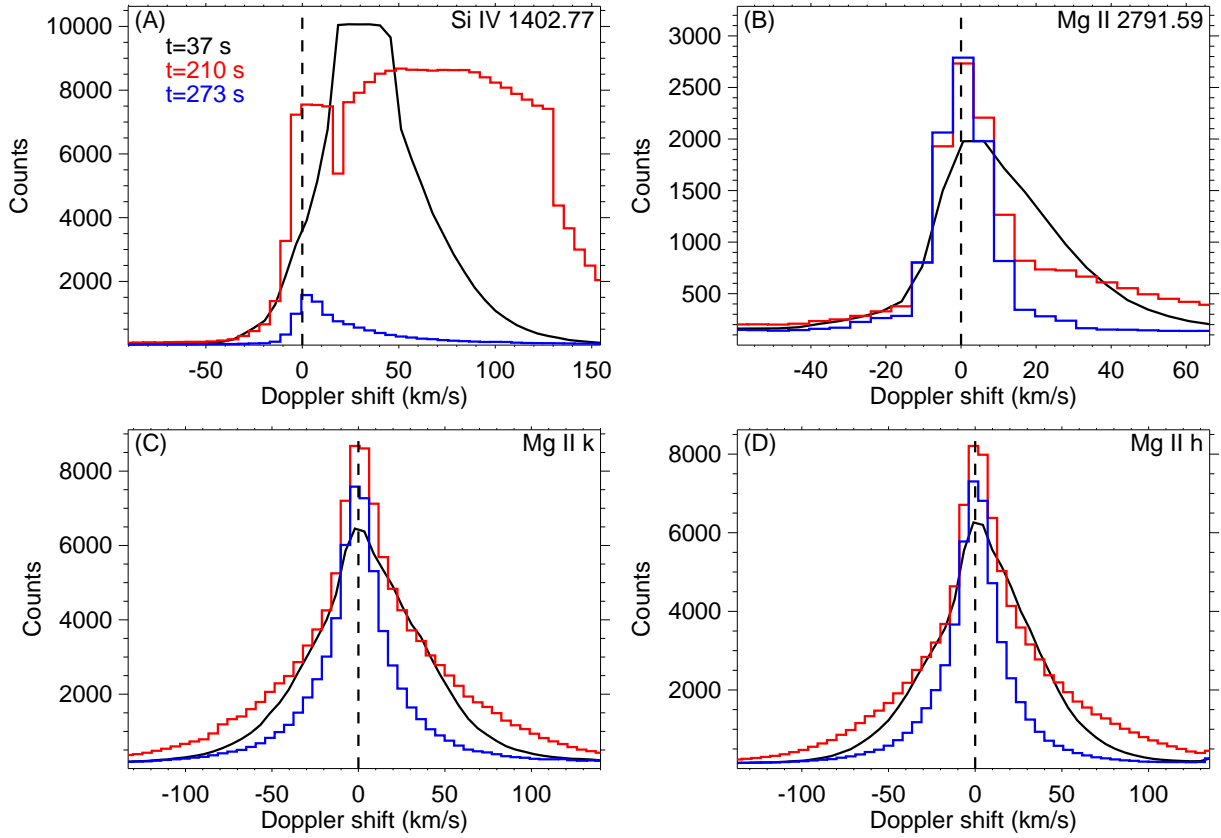}} \caption{ Spectral line profiles of Si~{\sc{iv}}~1402.77 \AA{}, Mg~{\sc{ii}}~2791.59 \AA{}, Mg~{\sc{ii}}~k and h at three different times in the northern ribbon. The times are shown in seconds after 11:41:32 UT (see Figure~\ref{fig.5}). The vertical dashed line in each panel indicates the rest wavelength of the corresponding line. } \label{fig.6}
\end{figure*}

\begin{figure*}
\centering {\includegraphics[width=\textwidth]{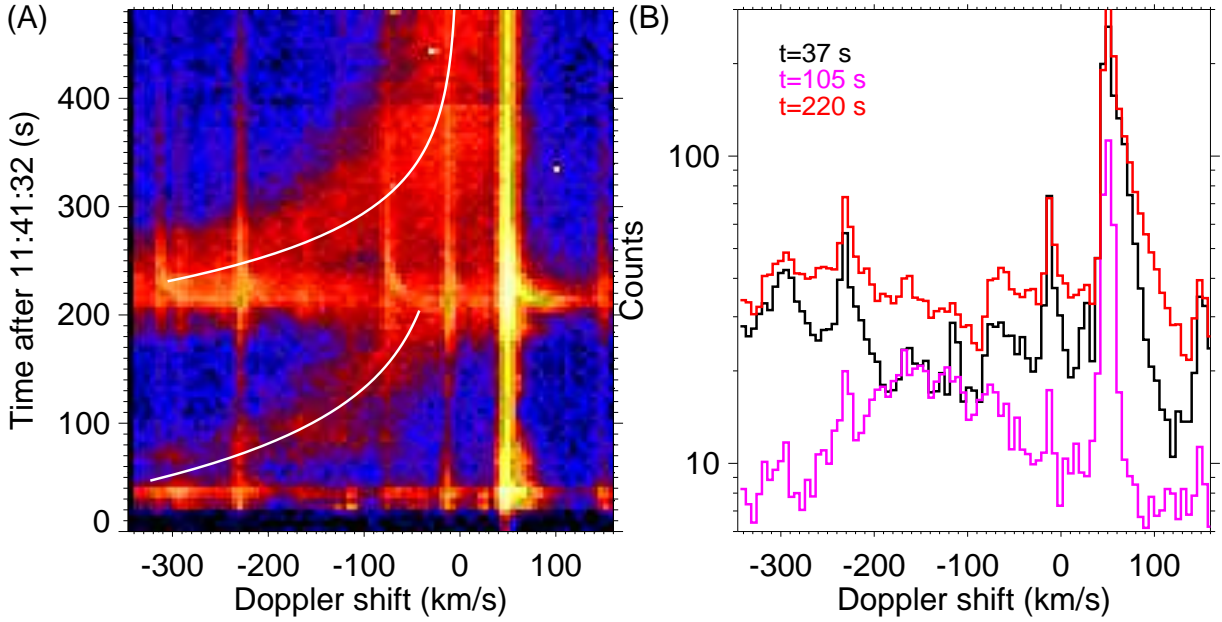}} \caption{ (A) Temporal evolution of the IRIS spectra in the 1354 \AA{}~spectral window at the northern ribbon. The two white curves mark two sessions of exponential decrease of the Fe~{\sc{xxi}}~blue shift. (B) The spectra at three different times (shown in seconds after 11:41:32 UT). } \label{fig.7}
\end{figure*}

\begin{figure*}
\centering {\includegraphics[width=0.6\textwidth]{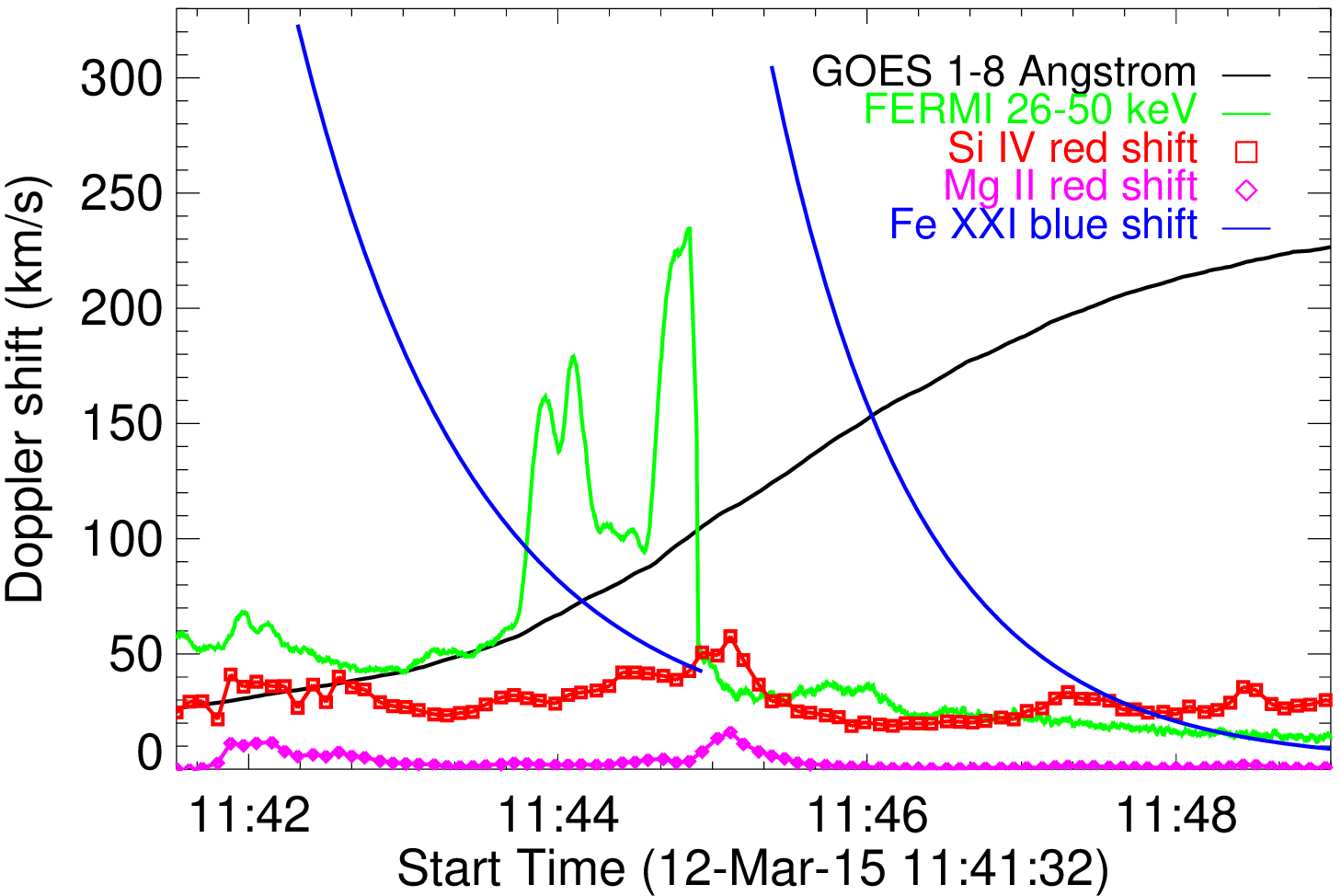}} \caption{ Time history of the GOES 1-8 \AA{} flux, FERMI 26--50 keV flux, Doppler shifts of Si~{\sc{iv}}~1402.77 \AA{}, Mg~{\sc{ii}}~2791.59 \AA{} and Fe~{\sc{xxi}}~1354 \AA{} at the northern ribbon. } \label{fig.8}
\end{figure*}

The southern ribbon was crossed by the IRIS slit after 11:55 UT, only in the decay phase of the flare, while the northern ribbon was sampled by the slit during all phases of the flare. Hence we focus on the dynamics of the northern ribbon here.  

Figure~\ref{fig.4} presents the temporal evolution of the line parameters and the Mg~{\sc{ii}}~k/h ratio at the northern ribbon. The Si~{\sc{iv}}~line is always redshifted in both the impulsive and decay phases, and the average red shift is $\sim$35~km~s$^{-1}$. The red shift is enhanced quasi-periodically on a time scale of 1--3 minutes. Before 11:58 UT each episode of redshift enhancement is accompanied by an obvious enhancement of the Si~{\sc{iv}}~line intensity. Such a correlation, also found by \cite{Zhang2016a} and \cite{Brosius2015}, vanishes after 11:58 UT. The Doppler shifts of the Mg~{\sc{ii}}~lines are much smaller compared to that of the Si~{\sc{iv}}~line. Despite that, the quasi-periodic enhancement is also clearly seen in the Doppler shift. The redshift enhancement is most significant around 11:42 UT and 11:45 UT. These two episodes of redshift enhancement are also accompanied by an obvious increase in the intensities of the two Mg~{\sc{ii}}~lines. Interestingly, the Mg~{\sc{ii}}~k/h ratio decreases from $\sim$1.2 to $\sim$1.1 during these two episodes. The SJI 1330 \AA{}~image sequence (the online video) clearly reveals two moving brightenings crossing the slit, suggesting sampling of the heated chromospheric plasma by IRIS at these two instants.  

The wavelength-time plots shown in Figure~\ref{fig.5} reveal detailed evolution of the profiles of several spectral lines at the northern ribbon. Previous observations have shown that hot emission and cool emission are sometimes spatially offset from each other at or around flare ribbons \citep[e.g.,][]{Young2015,Tian2015}. Considering this, we present the IRIS spectra in the 1354 \AA{}~spectral window at not only the ribbon, but also two other locations slightly to the north of the ribbon. The persistent and periodically enhanced red-wing asymmetry is clearly seen in the Si~{\sc{iv}}~line profiles. The first two episodes of enhanced red-wing asymmetry around t = 37 s and 210 s, which correspond to 11:42 UT and 11:45 UT, respectively, are most evident in the Si~{\sc{iv}}~and Mg~{\sc{ii}}~2791.59 \AA{}~lines. Notably, the Mg~{\sc{ii}}~2791.59 \AA{}~line is nearly symmetric at most times.

Several examples of line profiles are presented in Figure~\ref{fig.6}. At t = 37 s and 210 s, or in the two most significant heating events, large red shift or obvious red-wing enhancement can be clearly seen in the Si~{\sc{iv}}~and Mg~{\sc{ii}}~2791.59 \AA{}~line profiles, though the Si~{\sc{iv}}~line appears to be saturated. Slight red-wing enhancement is also observed in the Mg~{\sc{ii}}~k and h lines. In the absence of significant heating, e.g., at t=273 s, the Si~{\sc{iv}}~line is still enhanced at the red wing. However, the three Mg~{\sc{ii}}~lines all exhibit symmetric spectral profiles. 

The different behavior of the Si~{\sc{iv}}~and three Mg~{\sc{ii}}~lines suggests that we may use these lines to disentangle different physical processes in flares. The red shift of Si~{\sc{iv}} is usually believed to be a signature of chromospheric condensation during explosive chromospheric evaporation. However, flare models usually predict a condensation time scale of $\sim$1 minute or less \citep[e.g.,][]{Fisher1989,RubiodaCosta2015}, which is obviously not in line with IRIS observations of the persistent red shift. Recently \cite{Reep2016} proposed that energy release on many small-scale strands within the flare loop may be responsible for the persistent red shift. However, in this case one should also observe persistent red shift in the Mg~{\sc{ii}}~lines, which is not seen in our data. We think that the persistent red shift of Si~{\sc{iv}}~at the flare ribbon is likely caused by both chromospheric condensation and cooling of the heated plasma \citep{Tian2015}. The Mg~{\sc{ii}}~2791.59 \AA{}~emission line is formed as a result of a steep temperature increase in the lower chromosphere \citep{Pereira2015}, thus should be sensitive to the impulsive heating due to sudden energy release in flares. On the contrary, the emission of Mg~{\sc{ii}}~2791.59 \AA{}~might not be sensitive to cooling downflows in the absence of chromospheric heating. Thus, the Mg~{\sc{ii}}~2791.59 \AA{}~line should show short-duration red-wing enhancement only at the presence of impulsive heating, which is exactly what Figures~\ref{fig.4}--\ref{fig.6} show and is consistent with the prediction of flare models. This demonstrates the potential of the Mg~{\sc{ii}}~triplet lines in the diagnostics of heating process in flares. It is also worth noting that the red-wing asymmetry of the Si~{\sc{iv}}~and three Mg~{\sc{ii}}~lines is not consistent with most existing flare models, which usually predict entirely redshifted profiles of these low-temperature lines. Future modeling efforts similar to those of \cite{Ding2001}, \cite{Liu2015} and \cite{RubiodaCosta2017} are required to quantitatively evaluate how the nonthermal electrons impact the Mg~{\sc{ii}}~triplet line profiles. 

Quasi-periodic enhancement of Si~{\sc{iv}}~line parameters, especially the intensity and Doppler shift, has been previously detected by IRIS at the ribbons of a few flares. The periods are mostly in the range of 1--6 minutes \citep{LiD2015a,LiT2015,Brosius2015,Brosius2016,Brannon2015,Milligan2017}, though shorter periods of 32--42 s have also been found by \cite{Zhang2016a} in a circular-ribbon flare. Similar time scales are found in our observation (Figure~\ref{fig.4}). In addition, we find that the Mg~{\sc{ii}}~h, k and triplet lines also reveal a similar quasi-periodic behavior. This quasi-periodicity likely results from bursty reconnection, which may release energy sporadically in the corona. The release energy propagates downwards along flare loops quasi-periodically, most likely in the form of accelerated electrons \citep[e.g.,][]{Nishizuka2009}. The energy is then deposited in the chromosphere successively, leading to quasi-periodic enhancement of the Si~{\sc{iv}}~and Mg~{\sc{ii}}~line intensities and Doppler shifts. 

In previous observations of these multi-episode dynamics of cool lines in flares, the Fe~{\sc{xxi}}~line showed no quasi-periodic signature or even no emission, suggesting that the evaporation flows are not heated to 11 MK or that the radiative cooling time of the Fe~{\sc{xxi}}~emission is too long for any detectable fluctuation \citep{Brosius2016}. However, our observation demonstrates that these cool line dynamics can be synchronized with the hot Fe~{\sc{xxi}}~emission. Figure~\ref{fig.5}(A)-(C) clearly reveals multiple episodes of chromospheric evaporation, exhibiting as multi-episode Fe~{\sc{xxi}}~emission. The first two episodes are clearly associated with the two brightenings crossing the slit at 11:42 UT (around t = 37 s) and 11:45 UT (around t = 210 s), respectively. And they are accompanied by the strong chromospheric condensation as revealed by the significant red shifts of the Si~{\sc{iv}}~and Mg~{\sc{ii}}~lines. There appears to be no Fe~{\sc{xxi}}~emission  until the sudden enhancement of the FUV and NUV continua at about 11:42 UT. In the first episode, the Fe~{\sc{xxi}}~line is entirely blueshifted. An examples of this type of line profiles is presented in Figure~\ref{fig.7} (t=105 s). The blue shift of Fe~{\sc{xxi}}~smoothly decreases to nearly zero within $\sim$3 minutes, much slower than the fast decrease of the red shift of Mg~{\sc{ii}}~2791.59 \AA{}. A similar behavior was previously reported by \cite{Tian2015}. Such a smooth decrease of the blue shift has not been reproduced in most existing flare models. In the second episode, the Fe~{\sc{xxi}}~line profiles appear to reveal two components, one nearly at rest and the other blueshifted. This could be explained as a superposition of the newly evaporated hot plasma on the flare loop that is filled with hot materials through the first episode of evaporation in the line of sight direction. However, an unambiguous decomposition of these two possible components is difficult due to the blend of many lines such as Fe~{\sc{ii}}~1353.023 \AA{}, Si~{\sc{ii}}~1353.718 \AA{} and Fe~{\sc{ii}}~1354.013 \AA{} \citep[e.g.,][]{Tian2015,Young2015}. The line profiles around the times of maximum continuum enhancement are also examplified in Figure~\ref{fig.7} (t=37 s and t=220 s). These two profiles reveal a bump at velocities higher than $\sim$200 km~s$^{-1}$, suggesting the presence of highly blueshifted Fe~{\sc{xxi}}~emission. Since the emission of Fe~{\sc{xxi}}~line is mostly very weak and blended with many other lines \citep[e.g.,][]{Tian2015,Young2015}, a reliable Gaussian fitting can not be preformed for many of the line profiles. As demonstrated in \cite{Tian2015}, the time evolution of the Fe~{\sc{xxi}}~blue shift often can be well fitted with an exponential function. Thus, we simply overplot two exponential functions which fit the observed blueshifted Fe~{\sc{xxi}}~emission in Figure~\ref{fig.7}(A). These two exponential functions are used to approximate the Fe~{\sc{xxi}}~blue shift as a function of time in these two episodes. Figure~\ref{fig.5} reveals more episodes of Fe~{\sc{xxi}}~emission, especially at locations slightly offset from the northern ribbon. However, the Fe~{\sc{xxi}}~emission in these episodes is generally much weaker and not accompanied by significant enhancement of the Si~{\sc{iv}}~and Mg~{\sc{ii}}~emission, which may indicate the occurrence of gentle chromospheric evaporation. 

Figure~\ref{fig.8} shows the time history of the GOES 1-8 \AA{} flux, FERMI 26--50 keV flux, Doppler shifts of Si~{\sc{iv}}~1402.77 \AA{}, Mg~{\sc{ii}}~2791.59 \AA{} and Fe~{\sc{xxi}}~1354 \AA{} at the northern ribbon during the first two episodes. Both episodes of evaporation occur before the SXR peak. And they are clearly associated with the two episodes of HXR bursts, which supports the scenario of electron driven evaporation \citep{Tian2015,LiD2015b,LiD2017a}. The accompanied chromospheric condensation is characterized by the enhanced red shifts of the Si~{\sc{iv}} and Mg~{\sc{ii}}~lines around 11:42 UT and 11:45 UT. Complete evolution of chromospheric evaporation and the simultaneous condensation were previously reported for a few flares \citep{Brosius2003,Brosius2004,Brosius2003,Tian2015,Graham2015,LiD2015b}. However, only one episode of evaporation was observed in most of these observations. Using data taken by the Extreme-ultraviolet Imaging Spectrometer \citep[e.g.,][]{Culhane2007} on board the Hinode mission, \cite{Brosius2013} detected two episodes of chromospheric evaporation in the Fe~{\sc{xxiii}}~263.8 \AA{} line. However, a continuous velocity decrease of the blueshifted component was not observed in the second episode.

The electron density at the northern ribbon has been derived using the intensity ratio of the O~{\sc{iv}}~1401.156 \AA{}/1399.774 \AA{}~lines. These two lines have enough signal to allow a measurement of the intensity ratio at the northern ribbon at 11:42 UT and 11:45 UT. We find an intensity ratio of $\sim$2.0. According to CHIANTI v7.1 \citep{Landi2013}, this ratio is outside the density sensitive range of this O~{\sc{iv}}~line pair (5.83 -- 2.37, corresponding to a density range of log {\it N}$_{e}$/cm$^{-3}$ =10$^{9}$ -- 10$^{13}$) and suggests that the electron density is likely higher than 10$^{13}$ cm$^{-3}$. We noticed that a similar high density was obtained by \cite{Lee2017} for another flare ribbon.

\section{Flare loop top}

\begin{figure*}
\centering {\includegraphics[width=0.8\textwidth]{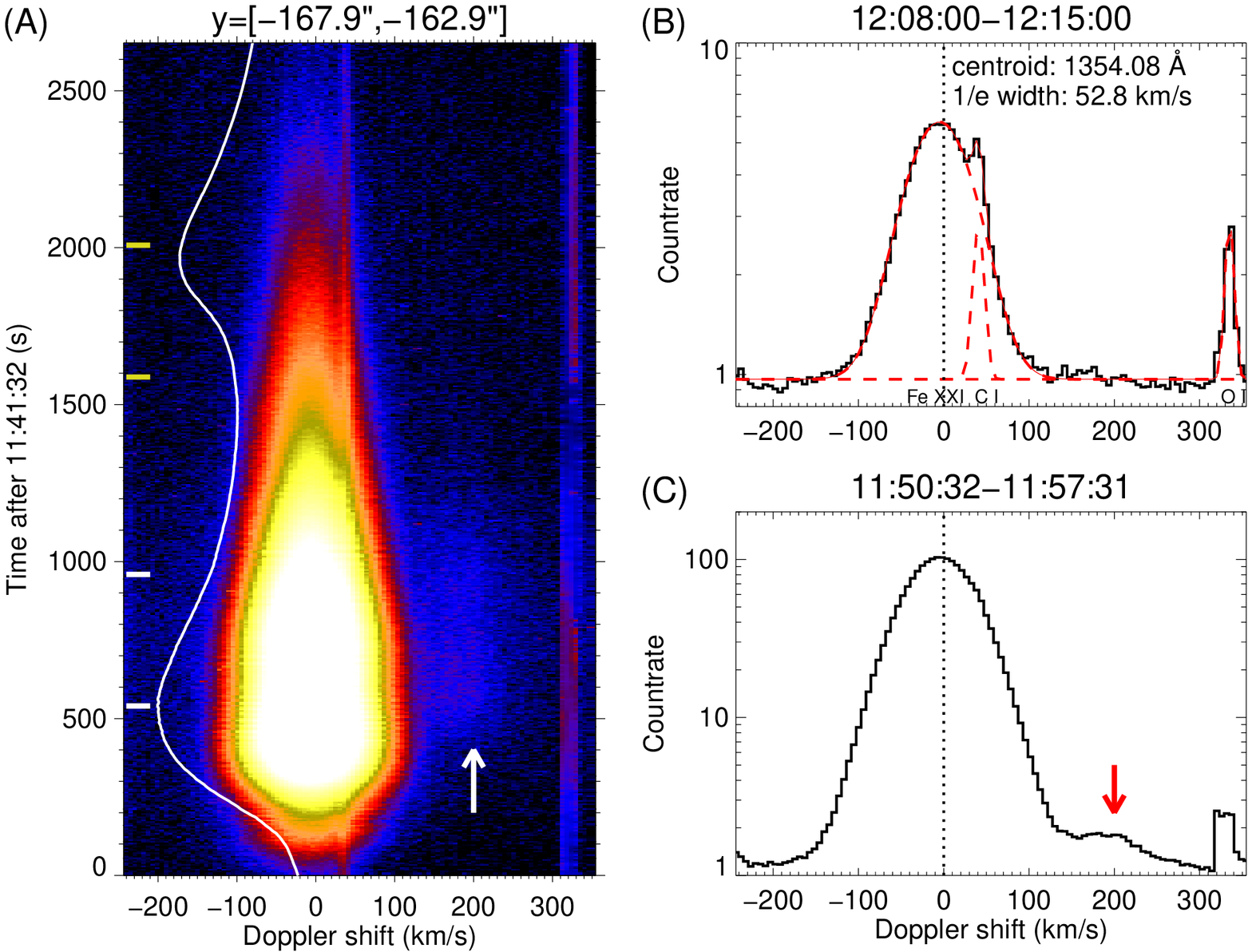}} \caption{ (A) Temporal evolution of the IRIS spectra in the 1354 \AA{}~spectral window at the loop top (averaged over solar-Y=[-167.9$^{\prime\prime}$,-162.9$^{\prime\prime}$]). (B) The line profile averaged in the time range of 12:08:00--12:15:00 UT (between the two yellow horizontal bars marked in (A)). The dashed lines represent a three-component Gaussian fit to the observed line profiles. The centroid position and exponential width of the Fe~{\sc{xxi}}~line are marked in this panel. (C) The line profile averaged in the time range of 11:50:32--11:57:31 UT (between the two white horizontal bars marked in (A)). The arrows indicate a redshifted component of the Fe~{\sc{xxi}}~line. } \label{fig.9}
\end{figure*}

Besides flare ribbons, the Fe~{\sc{xxi}}~1354 \AA{} line emission is also found at other locations such as post-flare loops  \citep{Young2015,Tian2016,LiD2017b} and pre-eruption flux ropes \citep{Zhou2016,Cheng2016}. Observations of the Fe~{\sc{xxi}}~1354 \AA{} line emission from flare loops may be used to estimate the rest wavelength of this forbidden line \citep{Young2015,Brosius2015}. We take the line profile averaged at the loop top (solar-Y=[-167.9$^{\prime\prime}$,-162.9$^{\prime\prime}$]) and during the period of 12:08:00--12:15:00 UT. This period is in the decay phase of the flare we study here. Note that the second SXR peak in the GOES light curve shown in Figure~\ref{fig.9}(A) was produced by a different flare. The average line profile is shown in Figure~\ref{fig.9}(B). Besides the Fe~{\sc{xxi}}~line, the C~{\sc{i}}~1354.288 \AA{} and O~{\sc{i}}~1355.5977 \AA{}~lines are also present in this spectral window. We have applied a triple Gaussian fit to the average line profile, and derived the line centroid and width of the Fe~{\sc{xxi}}~line. The line width (1/e width) is 52.8 km~s$^{-1}$. After subtracting the small instrumental broadening of $\sim$4 km~s$^{-1}$ \citep[see the discussion in Section S5 of the Supplementary Materials of][]{Tian2014b}, the line broadening is essentially only thermal. In this situation we may safely assume that the flare loops have a zero Doppler shift on average. Thus, the derived line centroid will be the rest wavelength of the Fe~{\sc{xxi}} line, which turns out to be 1354.0878$\pm$0.0072 \AA{}. The uncertainty is taken as the sigma value from the Gaussian fitting. This rest wavelength is very close to the one derived by \cite{Brosius2015}, and slightly smaller than the rest wavelength of 1354.106$\pm$0.023 \AA{} as derived by \cite{Young2015}. Note that here we have redone the wavelength calibration by assuming a zero Doppler shift for the C~{\sc{i}}~1354.288 \AA{} and O~{\sc{i}}~1355.5977 \AA{}~line profiles averaged over the same time period but outside the flare site (solar-Y=[-151$^{\prime\prime}$,-141$^{\prime\prime}$]). The derived Doppler shifts for these two cold lines are found to be redshifted by $\sim$2 km~s$^{-1}$, which likely reflects the uncertainty in the wavelength calibration. 

Figure~\ref{fig.9} also reveals an interesting feature of Fe~{\sc{xxi}}~emission at the loop top: a secondary component that is redshifted by $\sim$200 km~s$^{-1}$. High-speed downflows above the flare loop top were spectroscopically observed in only a few flares and they were often interpreted as downward moving reconnection outflows \citep{Wang2007,Hara2011,Tian2014,Simoes2015}. The highly redshifted emission component of Fe~{\sc{xxi}}~was detected after the impulsive phase, roughly in the time period of 11:48--12:00 UT. It is known that additional energy release by reconnection may continue in the decay phase of a flare. Thus, the observed secondary emission component might still come from the reconnection downflow. The density and/or temperature effect may explain the fact that this emission component was not observed in the impulsive phase. 

\section{Summary}
We have analyzed the IRIS observation of an M1.6 flare SOL2015-03-12T11:50. With a cadence of $\sim$5 seconds, this observation is among the highest-cadence spectroscopic observations of solar flares. The IRIS slit crossed both ribbons, and the northern ribbon was sampled by the slit during all phases of the flare. Our main results are summarized in the following.

(1) The ribbons generally reveal red shifts of the Si~{\sc{iv}}~and Mg~{\sc{ii}}~lines in both the impulsive phase and decay phase. The red shifts show quasi-periodic enhancement likely caused repeated chromospheric condensation in a scenario of bursty reconnection, especially at the northern ribbon. Before 11:58 UT each episode of redshift enhancement is also accompanied by an obvious enhancement of the Si~{\sc{iv}}~line intensity.

(2) The persistent and periodically enhanced red wing asymmetry of the Si~{\sc{iv}}~1402.77 \AA{} line suggests that this line responds to both cooling downflows and chromospheric condensation. However, the Mg~{\sc{ii}}~2791.59 \AA{} line appears to show short-duration (less than 1 minute) red-wing enhancement only at the times of impulsive heating, thus is mostly sensitive to the heating-related condensation process. 

(3) The first two episodes of redshift enhancement occurred around 11:42 UT and 11:45 UT, when two moving brightenings indicative of heating fronts were observed to cross the slit in the SJI 1330 \AA{} images. The greatly enhanced red wings of the Si~{\sc{iv}}~and Mg~{\sc{ii}}~lines are also accompanied by an obvious increase in the intensities of these lines and the HXR flux, suggesting multiple episodes of energy injection into the chromosphere in the form nonthermal electrons. 

(4) The Mg~{\sc{ii}}~k/h ratio has a small value of $\sim$1.2 at the ribbons. The ratio appears to be negatively correlated with the intensity of an Mg~{\sc{ii}}~triplet line or resonant line. The ratio decreases to $\sim$1.1 during the first two episodes.

(5) The first two episodes of red shift enhancement in the Si~{\sc{iv}}~and Mg~{\sc{ii}}~lines are also accompanied by two episodes of chromospheric evaporation. In the first episode, the Fe~{\sc{xxi}}~1354 \AA{}~line is entirely blueshifted. In the second episode, the Fe~{\sc{xxi}}~line reveals a nearly stationary component and a blueshifted component, which may result from a superposition of the newly evaporated hot plasma on the flare loop that is filled with hot materials through the first episode of evaporation in the line of sight direction. In both episodes, the blue shift smoothly decreases from $\sim$300 km~s$^{-1}$ to nearly zero within $\sim$3 minutes.

(6) More episodes of blueshifted Fe~{\sc{xxi}}~emission was found around the northern ribbon. However, the Fe~{\sc{xxi}} emission in these episodes is generally much weaker and not accompanied by significant enhancement of the Si~{\sc{iv}}~and Mg~{\sc{ii}}~emission, possibly indicating the occurrence of gentle chromospheric evaporation.

(7)  By assuming that the hot flare loops are stationary on average, we have derived a reference wavelength of the Fe~{\sc{xxi}} line, which is 1354.0878$\pm$0.0072 \AA{}. 

(8) At the flare loop top, the Fe~{\sc{xxi}} line reveals a secondary emission component that is redshifted by $\sim$200 km~s$^{-1}$, possibly indicating the downward moving reconnection outflow.

\begin{acknowledgements}
IRIS is a NASA Small Explorer mission developed and operated by LMSAL with mission operations executed at NASA Ames Research center and major contributions to downlink communications funded by ESA and the Norwegian Space Center. This work is supported by NSFC grants 11790304 and 11790300. N.C. is supported by the "Development of a Solar Coronagraph on International Space Station"  from the number of 2017185100. H.T. acknowledges support by ISSI/ISSI-BJ to the teams "Diagnosing heating mechanisms in solar flares through spectroscopic observations of flare ribbons" and "Pulsations in solar flares: matching observations and models". We thank Dr. P. R. Young and Prof. M.-D. Ding for helpful discussion.
\end{acknowledgements}

\end{document}